\documentclass[pdflatex,sn-mathphys-ay]{sn-jnl}

\usepackage[normalem]{ulem}
\usepackage{graphicx}%
\usepackage{multirow}%
\usepackage{amsmath,amssymb,amsfonts}%
\usepackage{amsthm}%
\usepackage{mathrsfs}%
\usepackage[title]{appendix}%
\usepackage{xcolor}%
\usepackage{textcomp}%
\usepackage{manyfoot}%
\usepackage{booktabs}%
\usepackage{listings}%
\usepackage{hyperref}
\usepackage{caption}
\usepackage{adjustbox}

\usepackage[linesnumbered]{algorithm2e} 
\RestyleAlgo{ruled} 

\usepackage{preamble}

\theoremstyle{thmstyleone}%
\theoremstyle{thmstyletwo}%

\theoremstyle{thmstylethree}%

\begin{document}

\title[Outlier detection in state-space models using mean-shift penalisation]{Outlier detection in state-space models using mean-shift penalisation}


\author*[1]{\fnm{Rajan} \sur{Shankar}}\email{rajan.shankar@sydney.edu.au}

\author[2]{\fnm{Ines} \sur{Wilms}}\email{i.wilms@maastrichtuniversity.nl}

\author[3]{\fnm{Jakob} \sur{Raymaekers}}\email{jakob.raymaekers@uantwerpen.be}

\author[1]{\fnm{Garth} \sur{Tarr}}\email{garth.tarr@sydney.edu.au}

\affil*[1]{\orgdiv{School of Mathematics and Statistics}, \orgname{The University of Sydney}, \orgaddress{\city{Sydney}, \postcode{2006}, \country{Australia}}}

\affil[2]{\orgdiv{Department of Quantitative Economics}, \orgname{Maastricht University}, \orgaddress{\city{Maastricht}, \postcode{6211LM}, \country{The Netherlands}}}

\affil[3]{\orgdiv{Department of Mathematics}, \orgname{University of Antwerp}, \orgaddress{\city{Antwerp}, \postcode{2020}, \country{Belgium}}}

\abstract{State-space models (SSMs) provide a flexible framework for modelling time series data, but their reliance on Gaussian error assumptions makes them highly sensitive to outliers. 
We propose a robust estimation method, ROAMS, that mitigates the influence of additive outliers by introducing shift parameters at each timepoint in the observation equation of the SSM. 
These parameters allow the model to attribute non-zero shifts to outliers while leaving clean observations unaffected. 
ROAMS then enables automatic outlier detection,
through the addition of a penalty term on the number of flagged outlying timepoints in the loss function, and simultaneous estimation of model parameters.
We apply the method to robustly estimate SSMs on both simulated data and real-world animal location-tracking data, demonstrating its ability to produce more reliable parameter estimates than classical methods and other benchmark methods. 
In addition to improved robustness, ROAMS offers practical diagnostic tools, including BIC curves for selecting tuning parameters and visualising outlier structure. 
These features make our approach broadly useful for researchers and practitioners working with contaminated time series data.}




\maketitle

\section{Introduction} \label{sec:introduction}

State-space models (SSMs) provide a powerful framework for dynamic analysis of time series across many domains, including engineering, econometrics, and ecology \citep{zhang_2017,sbrana_2023,auger-methe_guide_2021}.
SSMs separate latent system dynamics from observation noise and enable filtering, smoothing, and forecasting through Kalman-type recursions \citep{kalman_new_1960}. 
In practice, many real-world datasets are contaminated by outliers, and classical non-robust estimators of SSMs can be severely impacted. 
Even a small proportion of contamination can distort parameter estimates and forecasts. 
This is particularly problematic in applied settings where outliers are common, such as remote sensing, making robust estimation essential. 
In this paper, we address the challenge of joint parameter estimation and outlier identification in an SSM context.

Outlier detection and robust estimation in time series has a long history in the literature. 
Studies on detecting outliers in time series include, amongst others, \cite{fried2007rank, rousseeuw2019robust} for univariate time series, \cite{tsay_outliers_2000, galeano_outlier_2006, galeano2024detecting} for multivariate time series; see also the recent review by \cite{pena2023review} on high-dimensional time series. 
Robust estimation has been developed for classical time series models such as autoregressive moving average models (e.g., \citealp{bustos1986robust, chen1993joint, bianco2001outlier, muler2009robust}), their multivariate extensions (e.g., \citealp{ben1999robust, muler2013robust}), exponential smoothing models  (e.g., \citealp{croux2010robust, gelper2010robust}) and robust filtering methods e.g., \cite{fried2004robust, schettlinger2010real, borowski2014online}. 

For SSMs, the focus of this paper, outlier-robust variants of the Kalman filter \citep{kalman_new_1960} have been well studied, see amongst others \cite{cipra1997kalman, ruckdeschel_robust_2014, duran-martin_outlier-robust_2024,fang_robustifying_2018}.
Several robust estimation procedures for SSMs have been proposed, leveraging robust filters with a robustified likelihood (e.g., \citealp{crevits_robust_2019} for Huber maximum likelihood or maximum trimmed likelihood), assuming $t$-distributed observation processes (e.g., \citealp{ting_kalman_2007, agamennoni_outlier-robust_2011}), or using a generalised method of wavelet moments (e.g., \citealp{guerrier_robust_2022}).

Existing robust approaches to SSM typically only ameliorate the impact of outliers, for example using Huber-type likelihoods or heavy tailed data generating distributions, and do not necessarily provide explicit identification of outliers. 
Further, for trimming approaches, a priori knowledge about the level of contamination is required, for example, specifying the proportion of trimming required.  
None of the existing studies, however, propose a robust procedure where estimation of the SSM parameters and outlier detection are combined in one single loss function.

The method we propose in this paper fills this gap by providing an estimation framework that identifies additive outliers in the observation process and jointly delivers reliable parameter estimates in the presence of contamination. 
We introduce ROAMS, Robust Outlier-Adjusted Mean-Shift estimation for linear Gaussian SSMs. 
ROAMS augments the observation equation with a time-indexed mean-shift vector that negates the impact of outliers.
The challenge, though, is that these outlying timepoints are not known in advance, unlike regular missing observations, and need to be detected. 
ROAMS achieves this through the addition of an $L_0$ penalty on the shift parameters to the likelihood-based loss function.
Flagged observations with non-zero shifts are then excluded from state updates by setting the Kalman gain to zero, behaving as if they are missing observations.
Hence, the ROAMS procedure effectively leverages the ability of SSMs to accommodate missing observations.  
Achieving outlier robustness through mean shifts has been used successfully in the context of linear regression models \citep{she_outlier_2011,mccann_robust_2007,gannaz_robust_2007}, and more recently for covariance estimation \citep{raymaekers_mcd_2024} and Gaussian mixture models \citep{zaccaria_2025}.

In this paper we demonstrate that ROAMS provides a practical and effective framework for robust state-space modelling. 
Through an extensive set of simulation studies, we evaluate its performance under various contamination scenarios, showing that it reliably recovers model parameters, accurately identifies outliers, and offers an information criterion approach for data-driven tuning of the detection threshold. 
ROAMS is applied to real-world animal tracking data, including a blue whale example where we show a visual diagnostic that practitioners can use to select an appropriate tuning parameter. 
An open-source implementation, along with reproducible code for all simulations and applications, is publicly available at the GitHub page of the first author (\url{https://rajan-shankar.github.io/roams/}).

The remainder of the paper is structured as follows. 
Section \ref{sec:method} reviews the SSM framework and introduces our method, ROAMS, for outlier detection and robust estimation. 
Section \ref{sec:computational-aspects} describes the implementation, selection of the tuning parameter, and outlines how forecasts can be generated in an online setting. 
Section \ref{sec:simulations} presents three simulation studies exploring the performance under a range of contamination settings. 
Applications to real-world animal tracking data are given in Section \ref{sec:application}. 
Finally, Section \ref{sec:conclusion} summarises the key findings and outlines directions for future work.

\section{Method} \label{sec:method}

We begin by revising the linear Gaussian SSM in Section \ref{sec:lgssm}. In Section \ref{sec:outlier_mean_shift}, we present ROAMS --- our \textit{Robust Outlier-Adjusted Mean-Shift} method for robustly estimating SSMs.

\subsection{Linear Gaussian state-space model} \label{sec:lgssm}

We consider the linear Gaussian state-space model, 
\begin{align} \label{eq:SSM}
\begin{split}
    \by_t &= \bA\bx_t + \bv_t \\
    \bx_t &= \bPhi \bx_{t-1} + \bw_t \quad t=1,\ldots, n,
\end{split}
\end{align}
where $\by_t$ is the $p$-dimensional \textit{observation process}, $\bx_t$ is the hidden $q$-dimensional \textit{state process},  $\bv_t\sim N(\mathbf 0, \bSigma_\bv)$ is the \textit{observation error} and $\bw_t\sim N(\mathbf 0, \bSigma_\bw)$ is the \textit{state error}.  
The $p\times q$ matrix $\bA$ is the \textit{observation matrix} --- it captures how states are converted to observations and is typically treated as known, as we also do in the remainder of the paper. 
The $q \times q$ matrix $\bPhi$ is the \textit{state-transition matrix} --- it captures how states evolve over time and typically contains unknown parameters that need to be estimated. 

The observation process $\by_t$ is modelled by relying on the evolution of the state process $\bx_t$ that captures the underlying structure of the time series process.
To this end, the \textit{Kalman filter} is typically used as it provides minimum-variance unbiased estimates for $\bx_1,\dots,\bx_n$ under a linear Gaussian SSM via the \textit{Kalman recursions}.

Let $\bx_{t|s} := E(\bx_t|\by_1,\dots,\by_s)$ denote the conditional expected value of $\bx_t$ given the last $s$ observations and  $\bP_{t|s} := V(\bx_t|\by_1,\dots,\by_s)$ the conditional variance of $\bx_t$ given the last $s$ observations, with $s\le t$.
Furthermore, for initialisation purposes, let $\bx_{0|0}:=\bmu_0$ denote the initial state mean, and $\bP_{0|0}:=\bSigma_0$ the initial state variance.
The Kalman recursions are derived analytically (e.g., \citet{durbin2012time} or \citet[Chapter 6]{shumway2006time} for a textbook introduction) and given by
\begin{align} 
    \bx_{t|t-1} &= \bPhi \bx_{t-1|t-1} \nonumber \\
    \bP_{t|t-1} &= \bPhi \bP_{t-1|t-1} \bPhi^\top + \bSigma_\bw \nonumber \\
    \by_{t|t-1} &= \bA\bx_{t|t-1} \nonumber  \\
 \bS_{t|t-1} &= \bA \bP_{t|t-1} \bA^\top + \bSigma_\bv \nonumber  \\
      \bK_t &= \bP_{t|t-1} \bA^\top \bS_{t|t-1}^{-1} \label{eq:kalman_2} \\
    \bx_{t|t} &= \bx_{t|t-1} + \bK_t (\by_t - \by_{t|t-1}) \nonumber \\
    \bP_{t|t} &= \bP_{t|t-1} - \bK_t\bA\bP_{t|t-1}, \label{eq:kalman_4}
\end{align} 
where $\bK_t$ is referred to as the Kalman gain and $\bS_{t|t-1} := V(\by_t | \by_1,\dots,\by_{t-1})$.

These Kalman recursions  contain the unknown parameters $\bPhi, \bSigma_\bv, \bSigma_\bw$. We collect their elements into a parameter vector 
$\btheta$ and use Gaussian maximum likelihood to obtain estimates. In particular, we minimize the negative classical log-likelihood function of the observed data $\by_1,\dots,\by_n$ given by
\begin{align}
\begin{split}
-\ell_c(\btheta) = \frac{1}{2} \sum_{t=1}^n \left\{\log |\bS_{t|t-1}(\btheta)| + \text{MD}^2\left(\bvarepsilon_{t}(\btheta), \bS_{t|t-1}(\btheta)\right)\right\}, \label{eq:l_c}
\end{split}
\end{align}
where $\bvarepsilon_{t}(\btheta) = \by_t - \by_{t|t-1}(\btheta)$ denotes the error term at time $t$ and $\text{MD}(\bvarepsilon_t, \bS_t) = \sqrt{\bvarepsilon_t^\top\bS^{-1}_t\bvarepsilon_t}$ is the Mahalanobis distance. 
Note that the values $\by_{t|t-1}$ and $\bS_{t|t-1}$ computed from the Kalman filter have been written as functions of $\btheta$ to emphasise that they depend on $\btheta$. 
Throughout this paper, we use the \texttt{R} package \texttt{dlm} \citep{petris_2009, petris_2010} to estimate the parameters of the SSM.
In the remainder, we will refer to the resulting procedure as the classical estimator.

The classical estimator, however, is not robust to the presence of outliers. 
In the next section, we explain our robustification procedure.

\subsection{Outlier robustness via a  mean-shift model} \label{sec:outlier_mean_shift}

To robustly estimate the SSM in \eqref{eq:SSM}, we propose the \textit{Robust Outlier-Adjusted Mean-Shift} procedure, ROAMS in short.  
We address outliers in the observation process by introducing an additional $p$-dimensional parameter $\bgamma_t$ into the observation equation of \eqref{eq:SSM}. The SSM with mean-shift parameters $\bgamma_1,\dots,\bgamma_n$ is then given by 
\begin{align} \label{eq:mean_shift}
\begin{split}
    \by_t &= \bA \bx_t + \bgamma_t + \bv_t, \\
    \bx_t &= \bPhi \bx_{t-1} + \bw_t \quad t=1,\ldots, n.
\end{split}
\end{align}
For clean timepoints, the shift parameter $\bgamma_t$ should be $\mathbf 0$ so it has no influence on the observation equation. 
For outlying timepoints, $\bgamma_t$ should be non-zero, suggesting that the timepoint has been affected by an `external' shift. 
Such additive observational outliers have no influence on the subsequent observations; this in contrast to innovation outliers.\footnote{An innovation outlier could be modelled through a shift parameter in the state equation of the SSM. Then an outlying state $\bx_t$ would affect the current observation $\by_t$, but also subsequent states and therefore subsequent observations.  We leave the treatment of innovation outliers for future research.}

To jointly estimate $\btheta$ and the $\bgamma_t$ parameters in \eqref{eq:mean_shift} so that not all $\bgamma_t$ are trivially non-zero, we use a penalised approach, taking inspiration from \cite{she_outlier_2011} in the context of linear regression models.
Let $\bGamma_n := [\bgamma_1 \ \cdots \ \bgamma_n]$ denote the $p\times n$ matrix with shift parameters across all timepoints as column vectors. 
Also let the error $\bvarepsilon_t$ and  the observation variance $\bS_{t|t-1}$ depend not only on $\btheta$, but also on the previous $t-1$ shift parameters collected in $\bGamma_{t-1}$, which denotes the submatrix of $\bGamma_n$ containing the shift parameters from timepoint one until $t-1$ in its columns, and by convention let $\bGamma_0={\bf 0}$ for initialisation purposes. 

We propose the following penalised loss function,
\begin{align} \label{eq:penalised_loglik}
    -\ell_r(\btheta, \bGamma_n) + \frac{\lambda^2}{2} \sum_{t=1}^n 1_{\{\bgamma_t \neq \mathbf 0\}},
\end{align}
where 
\begin{align}
\small
    -\ell_r(\btheta, \bGamma_n) = \frac{1}{2}\sum_{t=1}^n \left\{1_{\{\bgamma_t = \mathbf 0\}} \log |\bS_{t|t-1}(\btheta,\bGamma_{t-1})| + \text{MD}^2\left(\bvarepsilon_{t}(\btheta, \bGamma_{t-1}) - \bgamma_t, \bS_{t|t-1}(\btheta, \bGamma_{t-1})\right) \right\}, \label{eq:l_r}
\end{align}
and the $L_0$ penalty term counts the number of timepoints flagged as outlying. 
The tuning parameter $\lambda$ controls the strength of the penalty.

The loss function in \eqref{eq:l_r} is equivalent to the negative log-likelihood of the SSM in \eqref{eq:mean_shift}, with two modifications to ensure that outlier influence is completely removed. The first modification is that we pre-multiply the term $\log|\bS_{t|t-1}|$ by the indicator function $1_{\{\bgamma_t = \mathbf 0\}}$, which equals one for clean timepoints (and zero otherwise). Without this indicator function, the presence of the $\log|\bS_{t|t-1}|$ term when $\by_t$ is outlying would unfairly penalise the estimates of the error variances $\bSigma_v$ and $\bSigma_w$. The second modification is that $\bvarepsilon_t$ and $\bS_{t|t-1}$ depend on $\bGamma_{t-1}$; we induce this dependence by setting the Kalman gain in equation \eqref{eq:kalman_2} to $\mathbf 0$ if $\by_t$ is an outlier, using
\begin{align} \label{eq:robust_kalman_gain}
    \bK_t = \bP_{t|t-1} \bA^\top \bS_{t|t-1}^{-1} \times 1_{\{\bgamma_t = \mathbf 0\}}.
\end{align}
The modified $\bK_t$ not only causes the state estimate $\bx_{t|t}$ to no longer use information from an outlying $\by_t$ when updating, but also causes the state variance $\bP_{t|t}$ to not decrease and instead remain unchanged: $\bP_{t|t} = \bP_{t|t-1}$. A comparison of how the second modification affects the standard Kalman recursions is presented in Table A1 in the supplementary material. In summary, both modifications to the negative log-likelihood of the SSM in \eqref{eq:mean_shift} use indicator functions to ensure the complete removal of the influence of outliers.

In fact, using indicator functions to ensure complete removal of the influence of outliers, \eqref{eq:l_r} corresponds to the \textit{observed-data} negative log-likelihood of the classical SSM as a function of $\btheta$, whereby the log-likelihood sums over only the `non-missing' observations, i.e. those where $\bgamma_t = \mathbf 0$. See \citet[Section 5.4]{Kitagawa1996} for a detailed discussion on log-likelihoods with missing observations in linear Gaussian SSMs.
Unlike regular missing observations, the outlying timepoints where $\bgamma_t \neq \mathbf 0$ are, however, not known in advance and need to be detected during the estimation procedure.

\section{Computational aspects}
\label{sec:computational-aspects}

Having introduced the ROAMS framework, we next discuss the computational strategies required for estimation. 
In Section \ref{sec:algorithm}, we detail the algorithm we use to solve the loss function in \eqref{eq:penalised_loglik}. 
In Section \ref{sec:lambda}, we discuss the BIC criterion to select the tuning parameter $\lambda$.
Section \ref{sec:online_state_est} provides additional details on the online state estimation, needed to estimate future states based on incoming data streams.
Finally, Section \ref{sec:software} ends with an overview of the \texttt{R} package \texttt{roams} that implements our procedure.

\subsection{Algorithm}
\label{sec:algorithm}

To minimise the robust loss function in \eqref{eq:penalised_loglik} over $\btheta$ and $\bGamma_n$, we use an iterative thresholding procedure.  In this section, we use superscripts in round brackets to denote the iteration, starting with the initialisation of $\bGamma_{n}^{(0)} = \mathbf 0$. At each iteration $k\ge1$, we first update $\btheta$ by minimising \eqref{eq:l_r} with fixed shift parameters $\bGamma_{n}^{(k-1)}$, as given by
\begin{align}
\btheta^{(k)} 
    &= \argmin_{\btheta} -\ell_r(\btheta, \bGamma_n^{(k-1)}).
    \label{eq:iterative_theta}
\end{align}
Then, we update the shift parameters, for timepoints $t=1, \ldots,n$, according to the \textit{hard-thresholding} rule,
\begin{align} \label{eq:thresholding_rule}
\begin{split}
    \bgamma_t^{(k)} = 
    \begin{cases} 
    \br_t^{(k)} & \text{if } \sqrt{\log\Big|\bS_{t|t-1}^{(k)}\Big| + \text{MD}^2\left(\br_t^{(k)}, \bS_{t|t-1}^{(k)}\right)} > \lambda,\\
        \mathbf 0 & \text{otherwise,}
    \end{cases}
\end{split}
\end{align}
with 
$\bS_{t|t-1}^{(k)} = \bS_{t|t-1}\left(\btheta^{(k)}, \bGamma_{t-1}^{(k-1)}\right)$ 
and residuals 
$\br_t^{(k)} = \by_t - \hat{\by}_{t|t-1}(\btheta^{(k)}, \bGamma_{t-1}^{(k-1)})$.
The values $\hat{\by}_{t|t-1}(\cdot)$ and $\bS_{t|t-1}(\cdot)$ are outputs from the Kalman recursions where the Kalman gain is computed using equation \eqref{eq:robust_kalman_gain}. 
The hard-thresholding rule minimises \eqref{eq:penalised_loglik} given $\btheta^{(k)}$ and $\bGamma_n^{(k-1)}$. Intuitively, the hard-thresholding rule says that if the Mahalanobis distance of a residual, after adjusting for some expected error variance through the term $\log|\bS_{t|t-1}(\cdot)|$, does not exceed $\lambda$, it is deemed ``clean" and retained ($ \bgamma_t^{(k)}  = {\bf 0}$).
By contrast, if it exceeds $\lambda$, then the corresponding timepoint is deemed ``extreme". In that case, setting $\bgamma_t^{(k)}$ equal to the residual will remove the influence of that timepoint on the next iteration of the estimation procedure.

Equations \eqref{eq:iterative_theta} and \eqref{eq:thresholding_rule} are iterated until convergence, as described in Algorithm \ref{alg:roams}. The procedure returns the final parameter estimates $\hat{\bGamma}_n$ and $\hat{\btheta}$. 
Each iteration amounts to fitting a classical SSM with missing observations and identifying outliers, and therefore has the same computational time complexity as a classical SSM fit. In practice, convergence is typically achieved within 10--20 iterations.

\subsection{Selecting the tuning parameter $\lambda$} \label{sec:lambda}
To select an appropriate tuning parameter $\lambda$, we propose to use the Bayesian information criterion (BIC) \citep{schwarz1978}.

We start by constructing a sequence of $J$ equally-spaced possible values from $\lambda_{\text{min}}$ to $\lambda_{\text{max}}$. 
From equation \eqref{eq:thresholding_rule}, $\lambda$ can be interpreted as a standardised threshold for determining whether or not an observation should be flagged as an outlier. 
As such, if $\lambda = 3$, then observations that are at least 3 Mahalanobis distance units away (after adjusting for expected error variance) from their predicted location should be flagged as outliers. 
This interpretation makes it easier to construct a sequence of candidate $\lambda$ values. 

Based on initial experimentation, we recommend starting the sequence at $\lambda_{\text{min}} = 2$ since too low values of $\lambda$ can result in too many timepoints being flagged as outlying; solutions flagging close to 50\% of the timepoints as outlying are unfavourable since outlier-robust methods are only sensible for contamination rates less than 50\%.

The largest value in the sequence, $\lambda_\text{max}$, is chosen to be the largest Mahalanobis residual from the fit using the classical estimate of $\btheta$, i.e. $\hat\btheta_{c}$ resulting from minimisation of \eqref{eq:l_c},
\begin{align*}
    \lambda_{\text{max}} = \max_{t=1,\dots,n} \text{MD}\left(\by_t - \hat{\by}_{t|t-1}(\hat\btheta_c), \bS_{t|t-1}(\hat\btheta_{c})\right).
\end{align*}
We then select the optimal $\lambda$ as the one that minimises
\begin{align*}
    \text{BIC}(\lambda) =  - 2 \ell_r\left(\hat\btheta_\lambda, \hat\bGamma_{n,\lambda}\right) + k_\lambda\log(n),
\end{align*}
where we explicitly denote the dependence of the estimates on $\lambda$, and $k_\lambda$ is the number of timepoints flagged as outliers, as given by the number of non-zero $\bgamma_{t,\lambda}$ ($t=1,\ldots,n)$.
The full tuning parameter selection procedure is built into Algorithm \ref{alg:roams}.

Finally, note that for some applications, practitioners may have a prior belief on what proportion of their data is outlying; let $\pi$ denote this proportion. 
In such settings, one could opt to select $\lambda$ such that $k_\lambda/ n \approx \pi$ instead of using the BIC.

\begin{algorithm}[H]
\caption{ROAMS (Robust Outlier-Adjusted Mean-Shift) estimation for linear Gaussian SSMs with tuning parameter selection}
\label{alg:roams}
\SetNlSty{}{}{:}
\DontPrintSemicolon
\KwIn{Observations $\by_1, \dots, \by_n \in \mathbb{R}^{p}$;
  SSM specification function $\texttt{build}(\btheta)$ returning $\bA, \bPhi, \bSigma_v, \bSigma_w, \bmu_0, \bSigma_0$;
  Initial parameter vector $\btheta^{(0)}$;
  Number of $\lambda$ values to consider $J\in\mathbb{N^+}$;
  Convergence tolerance $\varepsilon > 0$.
}
\KwOut{Selected $\lambda^\star$; Parameter estimate $\hat\btheta$; Mean-shift matrix $\hat\bGamma_n=[\hat\bgamma_1 \ \cdots \ \hat\bgamma_n]$.
}

\BlankLine
\tcp{\textbf{\textrm{Construct $\lambda$-sequence}}}
$\hat\btheta_{c} \leftarrow \argmin_\btheta \{-\ell_c(\btheta)\}$\; 

Run Kalman filter with $\hat\btheta_{c}$ to obtain $\hat\by_{t|t-1}$, $\bS_{t|t-1}$, and $\br_t = \by_t - \hat\by_{t|t-1}$ for $t=1,\dots,n$\;

Set $\lambda_{1} \leftarrow 2$, $\lambda_{J} \leftarrow \max_{t=1}^n \text{MD}(\br_t, \bS_{t|t-1})$, and construct an equally-spaced sequence $\{\lambda_j\}_{j=1}^J$\;

\BlankLine
\tcp{\textbf{\textrm{Fit model for each $\lambda_j$}}}
\For{$j=1,\dots,J$}{

  Set $k\leftarrow 1$, $\bgamma^{(0)}_t\leftarrow\mathbf 0_{p}$ for $t = 1,\dots,n$, and $\bGamma_n^{(0)} \leftarrow [\bgamma^{(0)}_1 \ \cdots \ \bgamma^{(0)}_n]$\;

  \Repeat{
    $\big\|\btheta^{(k)}-\btheta^{(k-1)}\big\|_\infty \le \varepsilon \;\wedge\; \big\|\bGamma_n^{(k)}-\bGamma_n^{(k-1)}\big\|_\infty \le \varepsilon$
  }{

    $\btheta^{(k)} \leftarrow \argmin_{\btheta} \{-\ell_r(\btheta, \bGamma_n^{(k-1)})\}$\;

    Run Kalman filter with $\btheta^{(k)}$, where Kalman gain $\bK_t = \mathbf 0_{p\times p}$ for timepoints $t : \bgamma^{(k-1)}_t \neq \mathbf 0_p$, to obtain $\hat \by_{t|t-1}$, $\bS_{t|t-1}$, and residuals $\br_t = \by_t - \hat \by_{t|t-1}$\;

    $\bgamma^{(k)}_t \leftarrow \begin{cases}
      \br_t & \text{if } \sqrt{\log|\bS_{t|t-1}| + \text{MD}^2(\br_t, \bS_{t|t-1})} > \lambda_j\\
      \mathbf 0_p & \text{otherwise}
    \end{cases}$ for $t=1,\ldots,n$\;

    \If{$\#\{t:\bgamma^{(k)}_t \neq \mathbf 0\} \ge \frac 1 2 n$}{
    Set $\bgamma^{(k)}_t\leftarrow\mathbf 0_p$ for $t$ with smallest value of $\sqrt{\log|\bS_{t|t-1}| + \text{MD}^2(\br_t, \bS_{t|t-1})}$ until $\#\{t:\bgamma^{(k)}_t \neq \mathbf 0\} < \frac 1 2 n$\;
    }
    
    $\bGamma_n^{(k)} \leftarrow [\bgamma^{(k)}_1 \ \cdots \ \bgamma^{(k)}_n]$\;

    $k \leftarrow k+1$\;
  }

  \BlankLine
  \tcp{\textbf{\textrm{Compute BIC at $\lambda_j$}}}
  Let $\hat\btheta_{\lambda_j} \leftarrow \btheta^{(k)}$, $\hat\bgamma_{t,\lambda_j} \leftarrow \bgamma_t^{(k-1)}$ for $t=1,\dots,n$, and $\hat\bGamma_{n,\lambda_j} \leftarrow [\hat\bgamma_{1,\lambda_j} \ \cdots \ \hat\bgamma_{n,\lambda_j}]$\;
  
  Let $k_{\lambda_j} \leftarrow \#\{t:\hat\bgamma_{t,\lambda_j} \neq \mathbf 0\}$\;
  
  $\text{BIC}(\lambda_j) \leftarrow k_{\lambda_j} \log(n) - 2 \ell_r(\hat\btheta_{\lambda_j}, \hat\bGamma_{n,\lambda_j})$\;
}

\BlankLine
\tcp{\textbf{\textrm{Extract minimum-BIC model}}}

Let $j^\star \leftarrow \argmin_{j=1}^J\text{BIC}(\lambda_j)$, 
$\lambda^\star \leftarrow \lambda_{j^\star}$, 
$\hat\btheta \leftarrow \hat\btheta_{\lambda^\star}$, 
and $\hat\bGamma_n \leftarrow \hat\bGamma_{n,\lambda^\star}$

\Return $\lambda^\star, \hat\btheta, \hat\bGamma_n$
\end{algorithm}

\subsection{Online state estimation} \label{sec:online_state_est}

Given parameter estimates  $\hat\btheta$, future states can be estimated based on an incoming stream of observations $\by_{n+1}, \by_{n+2}, \dots$; a process known as online state estimation. 
Online states are typically computed using a filter. 
A natural question to ask is which filter one should use.
If one suspects that future timepoints will be free of outliers, then the usual Kalman filter with parameter estimates $\hat\btheta$ will perform well. 
However, if outliers may appear in future timepoints, then a robust filter is needed. 
Below, we detail the robust filter we propose to use for this purpose.

ROAMS assumes that in-sample outlying timepoints are additive outliers in the observation equation of the SSM; they provide no information about their respective states and do not impact the process in future time periods. As such, ROAMS deals with these outliers in a `detect-and-reject' fashion: flagging potential outliers and ensuring these outliers have zero influence on the parameter estimates.
Assuming that future outliers also provide no information about their respective states, a \textit{detect-and-reject} filter should be used for online state estimation. 

One such filter is the \textit{threshold} filter explored by \cite{duran-martin_outlier-robust_2024,ting_kalman_2007}. The threshold filter is computed in the same way as the usual Kalman filter, except that the Kalman gain $\bK_t$ is set to $\mathbf 0$ if the Mahalanobis distance of $\by_t$ from its predicted value $\by_{t|t-1}$ exceeds a threshold $c$, namely, 
\begin{align} \label{eq:threshold_filter}
    \bK_t = \bP_{t|t-1} \bA^\top \bS_{t|t-1}^{-1} \times 1_{\{\text{MD}(\by_t - \by_{t|t-1}, \bS_{t|t-1}) \le c \}},
\end{align}
which shows close similarity to  \eqref{eq:robust_kalman_gain} regarding the in-sample treatment of outliers.

The threshold $c$ is typically set to the square-root of the 99\% quantile of the $\chi^2_p$ distribution. 
This choice implies that on average 1\% of clean observations will be falsely detected as an outlier. In our experience, this is, however, not the case. 
Once an observation is detected as outlying, it can trigger a \textit{cascade} of outliers to be detected because it fails to update the filtered state; this phenomenon has also been observed by \cite[section E.2.1]{duran-martin_outlier-robust_2024}.

To discourage such a cascade from occurring, we propose the \textit{fast-updating} threshold filter. It retains equation \eqref{eq:threshold_filter} of the original threshold filter, but it modifies the update equation for $\bP_{t|t}$ as follows,
\begin{align*} 
    \bP_{t|t} = \begin{cases}
        \bP_{t|t-1} - \bK_t\bA\bP_{t|t-1} & \text{if } \bK_t \neq \mathbf 0 \\
        b\bP_{t|t-1} & \text{if } \bK_t = \mathbf 0.
    \end{cases}
\end{align*}
In contrast to the standard  recursion in equation \eqref{eq:kalman_4}, the update reduces to $\bP_{t|t} = b\bP_{t|t-1}$ when an outlier has been detected at time $t$ ($\bK_t = \mathbf 0$). 
The scalar $b \ge 1$ controls how quickly $\bP_{t|t}$, and consequently $\bS_{t+1|t}$, grows after an outlier is detected, allowing the filter to revert to the true process. Note that for $b = 1$, the original threshold filter would be recovered.
We recommend taking $b=2$ based on empirical validation, since too large values of $b$ would no longer perform well when multiple outliers  occur consecutively. 
Indeed, after correctly detecting the first few timepoints as outliers, the filter will treat  subsequent outliers as clean timepoints, causing it to incorrectly detect the truly clean timepoints as outliers instead.
In Section \ref{sec:study3}, we investigate the performance of the robust fast-updating threshold filter, compared to the standard Kalman filter, on an out-of-sample test set where outliers are present.

\subsection{Software implementation} \label{sec:software}

To facilitate adoption of ROAMS, we provide an open-source \texttt{R} \citep{r_language} package \texttt{roams}, available on the GitHub page of the first author (\url{https://rajan-shankar.github.io/roams/}). 
The package is designed to provide researchers and practitioners with a straightforward interface for applying our penalised mean-shift framework for detecting outliers to linear Gaussian state-space models. 
We built our package on top of the \texttt{dlm} \texttt{R} package \citep{petris_2010}, which provides a flexible framework for specifying and fitting classical state-space models. 
Leveraging \texttt{dlm} ensures that our implementation inherits both numerical stability and computational efficiency, while allowing us to focus on extending the estimation framework to handle outlier detection and robust modelling.

The \texttt{roams} package implements all aspects of our methodology in Sections \ref{sec:method} and \ref{sec:computational-aspects}. In particular, the package provides functions to (i) fit classical state-space models via maximum likelihood estimation, (ii) estimate robust state-space models using the ROAMS method, (iii) select the tuning parameter $\lambda$ using either the BIC or a user-specified outlier contamination proportion, and (iv) perform online state estimation using either the standard Kalman filter or the proposed fast-updating threshold filter. Furthermore, robust benchmark methods, described at the start of Section \ref{sec:simulations}, are implemented to facilitate reproducibility of the simulation and empirical applications we conduct in Sections \ref{sec:simulations} and \ref{sec:application}.

\section{Simulations} \label{sec:simulations}

We conduct three different simulation studies. Study 1 (Section \ref{sec:study1}) evaluates ROAMS under different outlier configurations. Study 2 (Section \ref{sec:study2}) stress-tests ROAMS under increasing outlier contamination and decreasing outlier distance.
Study 3 (Section~\ref{sec:study3}) demonstrates the need for a robust filter when out-of-sample timepoints are contaminated. 

\paragraph{Data generating process.}

We simulate data from SSM \eqref{eq:SSM}
with a $p=2$-dimensional observation process and a $q=4$-dimensional state process. 
We set the parameters in accordance with the model used for animal tracking purposes described in Section~\ref{sec:application}.
In particular, the observation matrix and state-transition matrix are given by,
\begin{equation} \label{DSRM:A-and-phi}
\bA = 
\left[
\begin{matrix}
1&0&0&0 \\
0&1&0&0
\end{matrix}
\right], \text{ and } 
\bPhi = 
\left[
\begin{matrix}
1+\phi&0&-\phi&0 \\
0&1+\phi&0&-\phi \\
1&0&0&0 \\
0&1&0&0 \\
\end{matrix}
\right],
\end{equation}
respectively, with autoregressive parameter $\phi=0.8$, the observation error covariance matrix $\bSigma_{\bv}=\diag(\sigma^2_{\bv,1},\sigma^2_{\bv,2})$ and the state error covariance matrix $\bSigma_{\bw}=\diag(\sigma^2_{\bw,1}, \sigma^2_{\bw,2}, 0, 0)$. We set $\sigma^2_{\bv,j} = 0.4$ and $\sigma^2_{\bw,j} = 0.1$ for $j=1,2$.
The observation error variances are larger than the state error variances so that there is `low signal' in the data. 
The true initial state $\bx_0$ is set to $[0,0,0,0]$ so that the first observation $\by_1$ will occur near the coordinate $[0,0]$. 
Once the data has been generated, the timepoints are randomly contaminated with outliers, where the outlier generating mechanism is study-dependent; see Sections \ref{sec:study1} and \ref{sec:study2} for the specific mechanisms used.

\paragraph{Benchmarks.} 

We compare ROAMS, with $\lambda$ selected from $J = 20$ candidate $\lambda$ values using the BIC, to four benchmark methods.
The ``oracle" method is the maximum likelihood estimator that knows which points are outliers and treats them as missing. If ROAMS successfully flags the correct timepoints as outliers, then its parameter estimates will coincide with those of the oracle method.
The ``classical" method is the classical maximum likelihood estimator which minimises the loss function in \eqref{eq:l_c}.
The last two benchmarks are robust maximum likelihood procedures proposed by \cite{crevits_robust_2019}.
The Huber maximum likelihood method (``Huber") 
replaces the quadratic loss function in the classical likelihood with the Huber function,
$$
\rho(x) = \begin{cases} 
x^2 & \text{if } x < k \\
2kx - k^2 & \text{if } x \ge k,
\end{cases}
$$
where the constant $k$ is set to the 0.95 quantile of a $\chi_p$ distribution.  
The maximum trimmed likelihood method (``trimmed") discards a fraction $\alpha$ of the observations with the largest Mahalanobis distances from the likelihood. To be as generous as possible, we set $\alpha$ to the true proportion of outliers across the different simulations. 
Both the Huber and trimmed method use the robust Kalman filter of \cite{cipra1997kalman}.

\paragraph{Initialisation.} 

To estimate the SSM parameters for each method, the distribution of the initial state $\bx_0$ is specified as the point mass at $[\by_1,\by_1]$. 
This specification assumes that $\by_1$ is the true location of the object being tracked at time $t=1$, which is sound as long as $\by_1$ is not outlying. In our simulation studies, we therefore restrict outliers to only be allowed to occur at timepoints $t = 2, \dots, n$.

All methods are implemented using the L-BFGS-B optimisation routine \citep{byrd1995}, which permits box constraints on the elements of the parameter vector~$\btheta$.
For the parameter $\phi$, we set the lower bound to zero and the upper bound to one. For the variance parameters, we use $10^{-12}$ as lower bound, to ensure that the error variance matrices never degenerate to zero, and use $\infty$ as upper bound.

Another important consideration with estimating SSM parameters is the choice of initial values. 
We initialise the observation error variance parameters with the median absolute deviation of the differences of the observations along each coordinate axis,
$$
    \sigma^{2 \ \text{init}}_{\bv,j} = \text{MAD}^2\left(\left\{y_{j,t} - y_{j,t-1} : t=2,\dots,n\right\}\right) ,
$$
for $j=1,2$, where the scalars $y_{1,t}$ and $y_{2,t}$ are the two elements of the observation vector $\by_t$, for example, the longitude and latitude in an animal tracking application (see Section~\ref{sec:animal_tracking}). This choice of initial value robustly captures the spread between observations in the data. The same initial values are used for the 
state error variances. 
Finally, the parameter $\phi$ ranges between $0$ and $1$, so we initialise using the midpoint, $0.5$.

\paragraph{Performance metrics.}

For each study we evaluate the performance of the methods considered using one or more of the metrics outlined below.
\begin{itemize}
    \item \textbf{Outlier detection.} 
    The \textit{sensitivity}, or true positive rate, is the proportion of actual outliers that are correctly flagged by ROAMS. 
    If all timepoints are truly uncontaminated, meaning they contain no outliers, then sensitivity is not defined. 
    Lower sensitivity could indicate \textit{masking} effects --- larger outliers inhibiting a method's ability to detect smaller outliers. 
    The \textit{specificity}, or true negative rate, is the proportion of all uncontaminated timepoints that are correctly \textit{not} flagged by ROAMS. 
    The specificity can always be computed in our simulation studies since there will always be at least some uncontaminated timepoints. Lower specificity could indicate \textit{swamping} effects --- incorrectly labelling clean timepoints as outliers.
    \item \textbf{Parameter estimation.} 
    We compute the root-mean-squared error (RMSE), being the $L_2$ norm of the difference between the estimated and true parameter for three different groups of parameters: the scalar $\phi$, the vector of observation error variances  $\left[\sigma^2_{\bv, 1}, \sigma^2_{\bv, 2}\right]$, and the state error variances $\left[\sigma^2_{\bw, 1}, \sigma^2_{\bw, 2}\right]$.
    \item \textbf{Out-of-sample forecast performance}: we compute the one-step-ahead mean-squared forecast error, 
    $$
    \text{MSFE} = \frac {1}{n_\text{oos}} \sum_{t=n+1}^{n + n_\text{oos}} \left\| \by_t - \hat\by_{t|t-1}(\hat\btheta) \right\|^2_2,
    $$
with $n_\text{oos}$ being the number of out-of-sample timepoints. We take $n_\text{oos}=20$ clean out-of-sample timepoints, and investigate the effect of out-of-sample outliers in Section \ref{sec:study3}. 
To compute the one-step-ahead forecasts, we take the parameter estimates $\hat\btheta$ for each method, plug these into a filter and run the filter on the out-of-sample timepoints.
For the oracle and classical method, we use the Kalman filter; for the Huber and trimmed method, we use the filter of \cite{cipra1997kalman}; and for ROAMS, we use both the Kalman filter (ROAMS-Kalman) and the fast-updating threshold filter (ROAMS-FUT).
For all methods, we set the initial state mean such that the first forecast is perfect, i.e. $\hat\by_{n+1|n}(\hat\btheta) \equiv \by_{n+1}$, to ensure that all methods have a `good' starting point for computing out-of-sample forecasts. 
\end{itemize}

\subsection{Study 1: Different outlier configurations} \label{sec:study1}

\paragraph{Outlier configurations.} We investigate the performance of the methods across four different configurations: 
\begin{itemize}
  \item \textbf{Clean data}: no outliers. This configuration is used as a baseline to assess performance in the absence of outliers.
  \item \textbf{Fixed distance}: outliers are generated by pushing observations a distance of 5 units in a random direction. 
  This configuration represents a setting where timepoints have been perturbed by some fixed amount.
  \item \textbf{Multi-level}: outliers are generated by pushing observations a distance of 3, 5, or 7 units in a random direction with a roughly equal number of outliers for each distance. These outliers are labelled small, medium, and large outliers, respectively. 
  This configuration is designed to test the methods' ability to detect masked outliers: large outliers may appear outlying, but smaller outliers may go undetected.
  \item \textbf{Cluster}: outliers are generated by replacing observations with points from a cluster centered at the location [20, 20] (recall that the trajectory starts near $[0,0]$). This cluster of points is normally distributed with a diagonal variance matrix with 4 on the main diagonal.
\end{itemize}

\begin{figure}[t]
    \centering
    \includegraphics[width=1\linewidth]{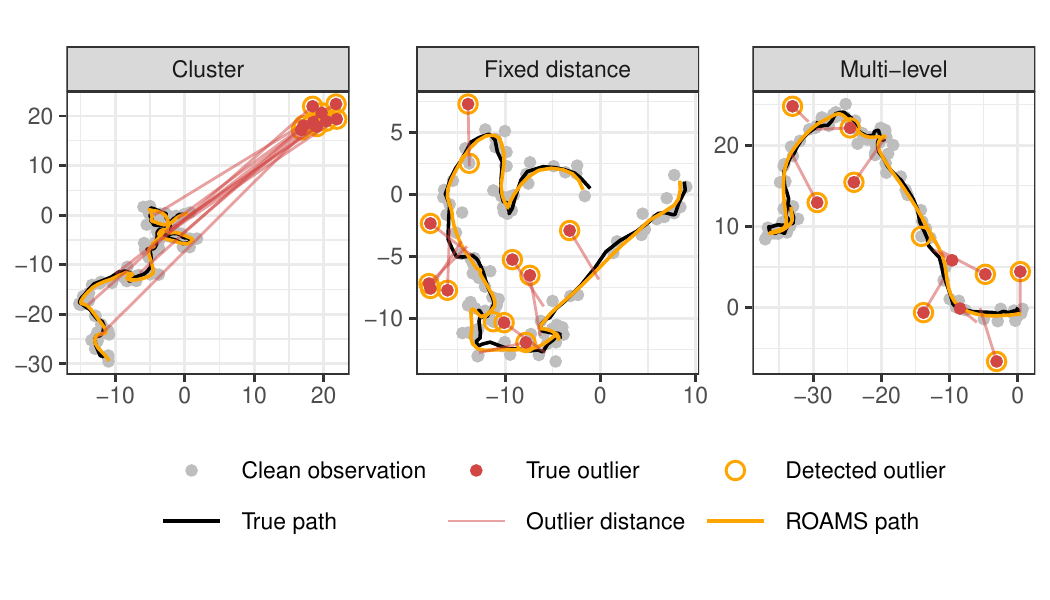}
    \caption{Three different outlier configurations, each with $n = 100$ timepoints and 10\% outliers.
    The true path is shown in black, along with clean and outlying observations. The red lines connect the outlying observations to their unobserved `clean' timepoints. The smoothed fit produced by ROAMS is shown by the orange path. The detected outliers are circled in orange.}
    \label{fig:datasets_study1}
\end{figure}

For all outlier configurations, we use a contamination rate of 10\% and randomly select the outlying timepoints (among $t=2, \ldots, n)$.
We use 200 simulation runs across all configurations and report results for different sample sizes, $n = 100, 200, 500, 1000$. 
An example data set for $n=100$ across each outlier configuration is displayed  in Figure \ref{fig:datasets_study1}. 

\paragraph{Results.}

\begin{figure}[t]
    \centering
    \includegraphics[width=1\linewidth]{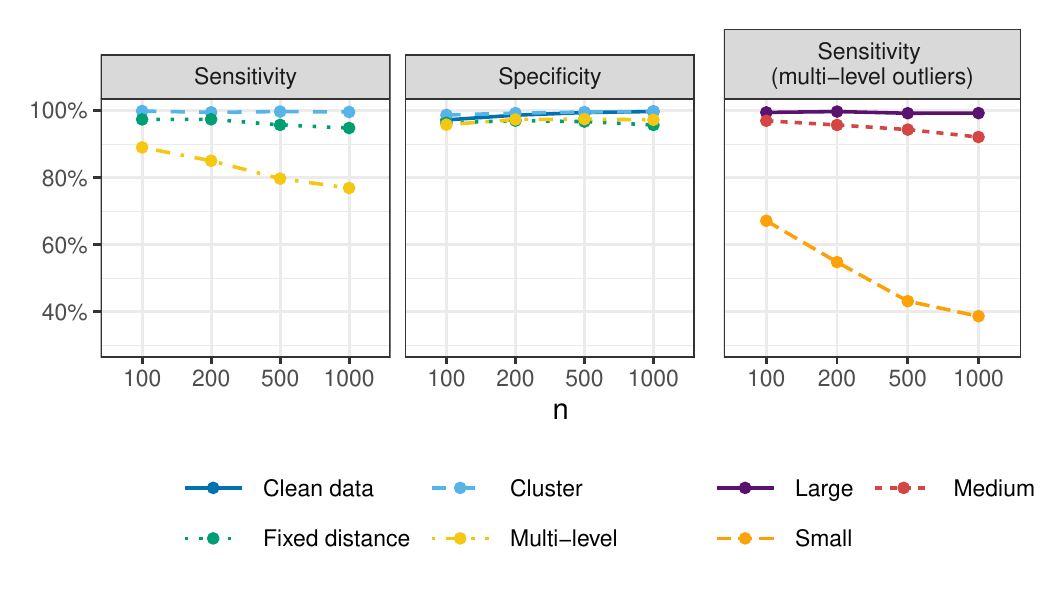}
    \caption{Sensitivity and specificity for ROAMS, averaged over all simulation runs and displayed for each sample size $n$ on the horizontal axis.
    \textit{Left, Middle}: Each line corresponds to a different configuration. \textit{Right}: The mean sensitivity of the multi-level configuration is broken down by outlier size: large, medium, and small.}
    \label{fig:sensspec_study1}
\end{figure}

First we discuss ROAMS's ability to correctly detect outliers.  
As seen in Figure \ref{fig:sensspec_study1}, the sensitivity of ROAMS in the fixed distance and cluster configurations is around 90--100\%. The cluster configuration in particular is very close to 100\% as its outliers are usually very easy to detect. Rarely, the true path goes \textit{through} the cluster of outliers at [20, 20]; in this circumstance, the sensitivity can drop to 80\% since it is then very difficult to detect true outliers when they sit close to clean timepoints.

The sensitivity of the fixed distance and multi-level outlier configurations slightly decreases as $n$ increases. From our empirical observation, ROAMS tends to over-flag timepoints as outliers for smaller $n$. Since less timepoints are flagged for larger $n$, the sensitivity decreases. An opposite effect can be seen with the specificity, which increases with $n$. Regardless of the sample size, ROAMS has the lowest sensitivity in the multi-level outlier configuration. This lower sensitivity compared to the other two outlier configurations can be attributed to the small outliers being hard to detect --- see the right-most panel of Figure \ref{fig:sensspec_study1}, which breaks down the multi-level sensitivity by outlier level: small, medium, and large. As expected, large outliers are most easily detected, whereas small outliers are only detected $\approx$ 40\% of the time at $n=1000$. The specificity shows similar results across all configurations: clean timepoints are correctly not flagged most of the time. For the clean data and cluster configurations, the specificity approaches 100\% as $n$ increases.

Table \ref{tab:par_RMSE_study1} displays the parameter accuracy in terms of the RMSEs, averaged across the simulation runs. 
In the clean data and cluster configurations, ROAMS has very similar RMSE to the oracle method across all parameters, and is competitive to the classical method on the clean data. 
In the cluster configuration, the Huber and trimmed methods have higher RMSEs than ROAMS since they cannot completely reject the cluster outliers, which are usually located far away from the rest of the data points. 
In the fixed distance and multi-level configurations, the Huber method is much more competitive, sometimes achieving a lower RMSE than ROAMS. 
These two configurations lend well to the Huber method because the outliers are much less extreme compared to those in the cluster configuration. 
The trimmed method tends to perform the worst out of the non-classical methods. 
For contaminated data, the classical method suffers, especially for the parameters $[\sigma^2_{\bv, 1}, \sigma^2_{\bv, 2}]$.
It gives all timepoints equal weighting during parameter estimation and thus becomes highly sensitive to outliers, which substantially inflate its estimates of observation error variance.

\begin{table*}
\caption{
RMSEs, averaged over the simulation runs, and reported for sample size $n = 1000$ for each parameter and configuration, across the five methods. The averaged RMSEs are reported relative to the oracle benchmark.}
\label{tab:par_RMSE_study1}
\centering
\begin{tabular}{clccccc}
\toprule
Parameter & Configuration & Oracle & ROAMS & Huber & Trimmed & Classical\\
\midrule
\multirow{4}{*}{$\phi$} & Clean data & 1.00 & 1.00 & 1.28 & 1.53 & 1.00\\
 & Cluster & 1.00 & 1.00 & 1.94 & 1.43 & 3.85\\
 & Fixed distance & 1.00 & 1.11 & 1.12 & 1.45 & 1.20\\
 & Multi-level & 1.00 & 1.13 & 1.07 & 1.49 & 1.16\\
\midrule
\multirow{4}{*}{$[\sigma^2_{\bv, 1}, \sigma^2_{\bv, 2}]$} & Clean data & 1.00 & 1.03 & 2.86 & 2.17 & 1.00\\
 & Cluster & 1.00 & 1.02 & 10.63 & 4.87 & 83.38\\
 & Fixed distance & 1.00 & 1.16 & 2.63 & 3.11 & 7.70\\
 & Multi-level & 1.00 & 1.71 & 2.50 & 2.83 & 7.99\\
\midrule
\multirow{4}{*}{$[\sigma^2_{\bw, 1}, \sigma^2_{\bw, 2}]$} & Clean data & 1.00 & 1.00 & 1.33 & 2.23 & 1.00\\
 & Cluster & 1.00 & 1.01 & 2.28 & 1.94 & 5.73\\
 & Fixed distance & 1.00 & 1.12 & 1.07 & 1.80 & 1.23\\
 & Multi-level & 1.00 & 1.18 & 1.10 & 1.91 & 1.21\\
\bottomrule
\end{tabular}
\end{table*}

\begin{figure}[t]
    \centering
    \includegraphics[width=1\linewidth]{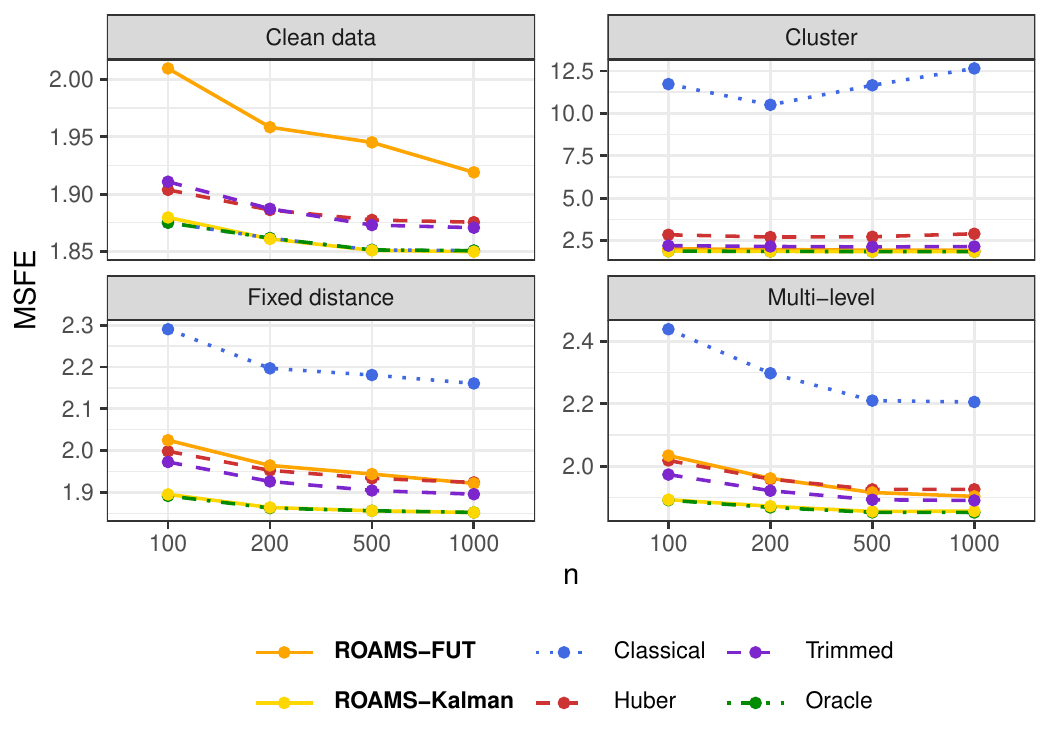}
    \caption{
    MSFE, averaged across the simulation runs,
    for different sample sizes $n$ (horizontal axis) and  configurations (panels).
    }
    \label{fig:MSFE_study1}
\end{figure}

Finally, Figure \ref{fig:MSFE_study1} reports out-of-sample performance on clean data. ROAMS-Kalman achieves very similar MSFE performance  to the oracle benchmark across all outlier configurations. 
In fact, when outliers are correctly detected, ROAMS-Kalman will have the \textit{same} performance as the oracle method since both methods use the Kalman filter out-of-sample.  
ROAMS-FUT performs worse, which is to be expected since it uses the robust fast-updating threshold filter and all out-of-sample points are clean. 

The Huber and trimmed methods provide some robustness to outliers, so they are much more accurate than the classical method. However, they cannot compete with ROAMS-Kalman due to ROAMS's ability to completely remove the influence of additive outliers. The Huber and trimmed methods outperform ROAMS-FUT on the fixed distance and multi-level configurations due to their ability to provide robustness against less-extreme outliers, but ROAMS-FUT performs better on the cluster configuration where outliers are usually more extreme, again due to ROAMS's ability to completely reject outliers. 
The classical method, as expected, shows poor out-of-sample performance on the outlier configurations. In the clean configuration, it coincides with the oracle method since the latter does not flag any outliers. For all methods, the MSFE decreases as the sample size increases up to $n=500$, after which the MSFE plateaus.

\subsection{Study 2: Increasing contamination rate and decreasing distance} \label{sec:study2}

We now investigate ROAMS's performance under increasingly challenging outlier settings, namely when the contamination proportion increases and/or the outliers become less extreme, making them harder to detect. 
To this end, we use the ``fixed distance" configuration from Study 1 with $n = 200$ and consider the following settings. 
\begin{itemize}
    \item \textbf{Contamination rate}: 0\%, 5\%, 10\%, 15\% and 20\%. The outlier distance is fixed at 5 units. 
    \item \textbf{Outlier distance}: 1, 3, 5, 7 and 9 units away from where the clean timepoint would have been, at a random angle. The contamination rate is fixed at 10\%. 
\end{itemize}
For each setting, we use 200 simulation runs. Note that we use the same clean data across all settings and add in outliers by replacing clean timepoints; the outliers are thus nested for increasing contamination rates.

\begin{figure}[t]
    \centering
    \includegraphics[width=1\linewidth]{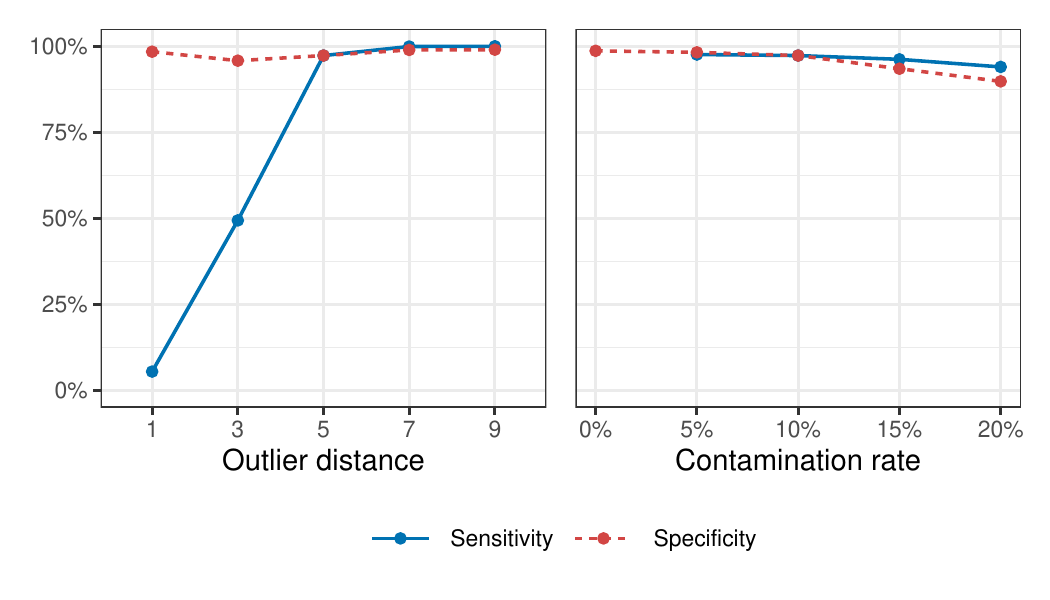}
    \caption{Sensitivity and specificity, averaged across the simulation runs, for increasing outlier distances (left) and contamination rates (right).}
    \label{fig:sensspec_study2}
\end{figure}

\paragraph{Results.}
Figure \ref{fig:sensspec_study2} illustrates ROAMS's ability to correctly classify outliers. 
At distances of 5, 7 and 9 units, outliers are easy to distinguish from clean timepoints, resulting in high sensitivity and specificity. 
Outliers of 1 and 3 units are difficult to detect, resulting in lower sensitivity. 
This difficulty was seen earlier in Study 1 Figure \ref{fig:sensspec_study1}, where the small multi-level outliers (also 3 units in distance) were hard to detect. 
Interestingly, since 1-unit outliers are even harder to detect, ROAMS tends to think most timepoints are clean, resulting in higher specificity than the 3-unit outliers case. 
As the contamination rate increases, the sensitivity and specificity degrade slightly, but it tells us that ROAMS can handle data sets with moderately high contamination rates.

We also compared ROAMS with $\lambda$ selected using the BIC to ROAMS where $\lambda$ is selected to flag a proportion of timepoints that is close to the true contamination rate, as described at the end of Section \ref{sec:lambda}. When the true contamination rate is not captured by any of the 20 candidate $\lambda$ values, we choose the $\lambda$ that is \textit{closest} to the true rate, as measured by the absolute difference from the true rate. In the case of a tie between over- and under-detection, the $\lambda$ with the higher detection rate is selected.
From Figure A1 in the supplementary material, the outlier detection performance between both selection approaches is very similar, but the true-contamination-rate ROAMS was slightly less sensitive and slightly more specific than the BIC ROAMS.

\begin{figure}[t]
    \centering
    \includegraphics[width=1\linewidth]{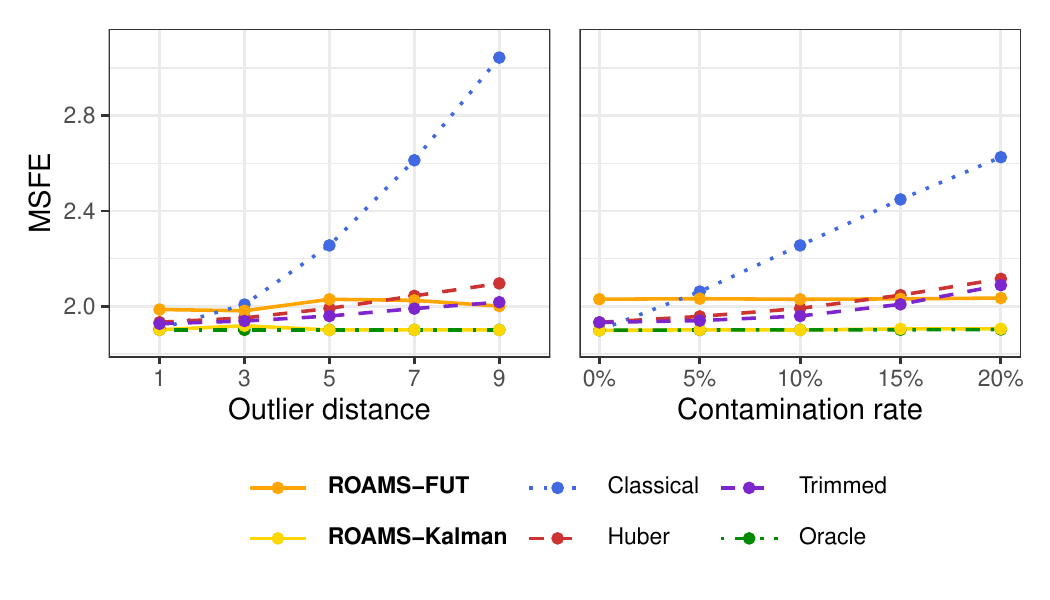}
    \caption{
    MSFE, averaged across the simulation runs, 
    for increasing outlier distances (left) and contamination rates (right).
    }
    \label{fig:MSFE_study2}
\end{figure}

We evaluate the performance of the methods using the out-of-sample MSFE in Figure \ref{fig:MSFE_study2}. 
Increasing the outlier distance or increasing the contamination rate hurts the classical method's MSFE, as expected. 
ROAMS-Kalman performs similarly to the oracle benchmark in both plots, except for when the outlier distance is 3 units where it performs slightly worse. 
As seen in Figure \ref{fig:sensspec_study2}, outliers of 3 units are difficult to detect, so ROAMS sometimes incorrectly includes them in the estimation process, which inflates the observation error variance parameters. 
The 1-unit outliers are even harder to detect, but since these outliers are so small, parameter estimation is not affected as much, resulting in better out-of-sample performance. 
Next, ROAMS-FUT performs worse than ROAMS-Kalman but its performance does not degrade as the outlier distance decreases or the contamination rate increases. 
The Huber method performs worse than its trimmed counterpart in both plots, likely because the trimmed method knows how many in-sample outliers to expect. 
The results of this study thus show that ROAMS is stable and can withstand outliers of a variety of different sizes and quantities.

\subsection{Study 3: Contamination in the testing set}
\label{sec:study3}

The simulated data in Study 2 (Section \ref{sec:study2}) did not contain contamination out-of-sample.
In many real-world applications, the out-of-sample data may not be free of outliers, hence we explore how the methods perform under such a scenario. 

We take the simulated data from Study 2, where the outlier distance is fixed at 5 units. 
For a given contamination rate, we incorporate the same rate of outliers in the out-of-sample data ($n_\text{oos} = 20$). 
The out-of-sample timepoints that are made outlying are chosen as follows:
\begin{itemize}
    \item \textbf{0\%}: Out-of-sample data remains as-is, hence clean.
    \item \textbf{5\%}: A single outlier at the 10th timepoint.
    \item \textbf{10\%}: 2 outliers at the 5th and 15th timepoints.
    \item \textbf{15\%}: 3 outliers at the 5th, 10th and 15th timepoints.
    \item \textbf{20\%}: 4 outliers at the 5th, 10th, 15th and 20th timepoints.
\end{itemize}
This fixed placement of outliers ensures there are no outlying observations at initialisation and prevents consecutive outliers. Controlling their spacing makes the results clearer and more interpretable.

To evaluate the performance of the methods, we again use the one-step-ahead MSFE, but we compute the forecast error based on the \textit{clean} observed timepoints (i.e. what the observed timepoints would be prior to adding in the outliers).

\begin{figure}[t]
    \centering
    \includegraphics[width=0.7\linewidth]{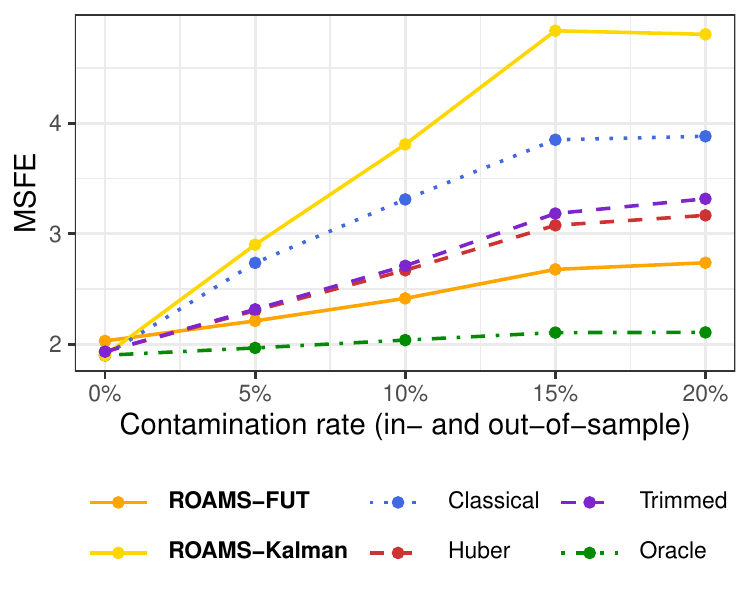}
    \caption{MSFE, averaged across the simulation runs, for increasing outlier contamination rates.
    }
    \label{fig:MSFE_study3}
\end{figure}

\paragraph{Results.}
The results of this study are shown in Figure \ref{fig:MSFE_study3}. 
The oracle method performs the best since outlying timepoints are treated as missing observations in its Kalman filter.
After the oracle method, ROAMS-FUT performs the best due to its ability to reject outliers based on its threshold filter. 
The two robust benchmarks, Huber and trimmed, perform worse than ROAMS-FUT but better than the classical method. 
Finally, ROAMS-Kalman performs worse than the classical method. 
The classical estimates of the observation error variance parameters are much higher than that of ROAMS --- $[2.28, 2.23]$ compared to ROAMS's $[0.38, 0.39]$ at the 15\% contamination setting, averaged across the simulation runs --- suggesting that the classical method does not `trust' observations as much. 
Therefore, when paired with the Kalman filter for out-of-sample forecasting, the classical method is not drawn to outliers (or any point for that matter) as much. 
On the other hand, the ROAMS method when paired with the Kalman filter, trusts that timepoints are reliable, due to ROAMS's lower estimates of the observation error variance parameters.
This `trust' backfires for ROAMS-Kalman as it gets drawn to outliers more easily, resulting in worse performance than the classical method. Hence, there is a need for a robust filter, such as the fast-updating threshold filter, to account for potential outliers in future timepoints.

\section{Animal tracking application} \label{sec:application}

We now demonstrate the usefulness of ROAMS for animal tracking applications.
In Section \ref{sec:animal_tracking}, we compare its performance to the classical method across four different animals. 
In Section \ref{sec:blue_whale}, we focus on the location tracking of a blue whale over time. 

\subsection{Animal tracking using the DCRW model} \label{sec:animal_tracking}

We showcase our method's ability to estimate an animal tracking model on four different animal tracking data sets: feral cat \citep{cat_data, roshier_2021}, polar bear \citep{polar_bear_data, auger-methe_guide_2021}, seal \citep{seal_data_package} and blue whale \citep{blue_whales_data, bailey_behavioural_2009}. These data sets contain the longitude and latitude measurements of a given animal over a period of time, obtained from satellite-based systems. Some processing has been performed on the data sets to ensure that the measurements occur at regular time intervals. 
For each data set, we divide it into a training set and a testing set using a 90\%/10\% train/test split. 
A summary of sample sizes, missing observations, and frequency of observations can be found in Table \ref{tab:animal_data_summary}. 

\begin{table*}
\caption{Sample size and frequency of observations for each data set, along with number of outliers detected by ROAMS (for training set) and ROAMS-FUT (for testing set). The column $n_\text{complete}$ is the number of non-missing observations.}
\label{tab:animal_data_summary}
\centering
\begin{tabular}{llcccccc}
\toprule
& & \multicolumn{3}{c}{Training set} & \multicolumn{3}{c}{Testing set} \\ \cmidrule(lr){3-5}\cmidrule(lr){6-8}
Data set & Frequency & $n$ & $n_\text{complete}$ & 
Detected & $n$ & $n_\text{complete}$ & Detected \\
\midrule
Cat & 20 min & 308 & 224 & 1 & 34 & 25 & 3\\
Polar bear & 1 day & 329 & 304 & 12 & 37 & 28 & 1\\
Seal & 1 day & 211 & 190 & 1 & 23 & 8 & 1\\
Blue whale & 12 hours & 110 & 90 & 11 & 12 & 11 & 1\\
\bottomrule
\end{tabular}
\end{table*}

We model the location of the animals using the first-differenced correlated random walk (DCRW) model, a linear Gaussian SSM that has been recently discussed by \cite{auger-methe_guide_2021} for its good performance in animal tracking. The DCRW model has the following observation and state processes,
\begin{align} \label{eq:DCRW_equations}
\begin{split}
    \by_t &= \bz_{t} + \bv_t \\
    \bz_t &= \bz_{t-1} + \phi(\bz_{t-1} - \bz_{t-2}) + \bw_t,
\end{split}
\end{align}
where $\by_t$ represents the location observed by the satellite, as captured by the longitude and latitude coordinates (hence $p=2$), 
$\bv_t$ is the error from the satellite measurements, $\bz_t$ represents the true location of the animal (also in terms of longitude and latitude), and $\bw_t$ captures the randomness in the animal's movement. 
The DCRW model assumes that the animal's true location depends on the previous true location $\bz_{t-1}$, and the previous velocity captured by the term 
($\bz_{t-1} - \bz_{t-2}$). The parameter $\phi\in [0,1)$ represents the autoregressive dependence. Note that model \eqref{eq:DCRW_equations} is not yet in the state-space form of equation \eqref{eq:SSM}; hence the usage of $\bz_t$ instead of $\bx_t$. Nonetheless, it can be equivalently represented in state-space form \eqref{eq:SSM} thereby using $q=4$ dimensional state process $\bx_t$ that contains the true (longitude and latitude) locations of the animal at times $t$ and $t-1$,   the observation matrix and state-transition matrix as given in equation \eqref{DSRM:A-and-phi}, and the observation error covariance matrix as well as the state error covariance matrix are assumed to be diagonal.

We apply both ROAMS and the classical method to the four animal tracking training sets.
The tuning parameter $\lambda$ for ROAMS is selected by BIC over a grid of $J = 50$ candidate values. A finer grid is used for the animal tracking data sets than for the simulated data in Section \ref{sec:simulations}, since the simulation studies require repeated model fitting across many runs.

To estimate the DCRW model, all initialisation and estimation choices are as discussed in Section \ref{sec:simulations}.
On the testing set, we compute one-step-ahead MSFEs using the Kalman filter with the classical estimates, and the Kalman and FUT filters for the ROAMS estimates. 
As there may be outliers in the testing set, the standard MSFE metric may not be reliable, so we compute a second type of MSFE: the `$\text{MSFE}_\text{clean}$', which removes forecasting errors associated with points detected as outliers by the ROAMS-FUT filter from the MSFE calculation. For fair comparison, the same points are also removed when computing the  $\text{MSFE}_\text{clean}$ for the classical method.

Finally, note that all data sets contain some missing observations. 
Missing observations are not an issue for model fitting nor for forecasting on the testing set, as we set $\bK_t = \mathbf 0$ for the missing timepoints. 
However, they do pose a problem for initialising the out-of-sample filters. 
We therefore use the first \textit{non-missing} observation in the testing set to initialise the filters.

\paragraph{Results.}

Table \ref{tab:animal_data_summary} shows the number of outliers detected by ROAMS in the training sets, and  the number of outliers detected by the out-of-sample filter ROAMS-FUT in the testing sets. Together with the plots of the animal trajectories (which can be found in one of the vignettes in the \texttt{roams} package), we note that the cat data set is fairly clean, while the polar bear, seal and whale data sets contain various amounts of outliers.

\begin{table*}[t]
\caption{Performances of classical and the two ROAMS filters on each data set. The $\phi$ estimates are reported together with the MSFE and $\text{MSFE}_\text{clean}$, which are reported relative to the classical method. Note that ROAMS-FUT and ROAMS-Kalman both use the ROAMS procedure on the training set, so they share the same $\hat\phi$.}
\label{tab:animal_data_performance}
\centering
\begin{tabular}{llccc}
\toprule
Data set & Method & $\hat\phi$ & MSFE & $\text{MSFE}_\text{clean}$\\
\midrule
\multirow{3}{*}{Cat} & Classical & 0.35 & 1.00 & 1.00\\
 & ROAMS-FUT & \multirow{2}{*}{0.44} & 1.70 & 1.69\\
 & ROAMS-Kalman &  & 1.04 & 1.04\\
\midrule
\multirow{3}{*}{Polar bear} & Classical & 0.61 & 1.00 & 1.00\\
 & ROAMS-FUT & \multirow{2}{*}{0.61} & 0.45 & 0.01\\
 & ROAMS-Kalman &  & 1.12 & 1.21\\
\midrule
\multirow{3}{*}{Seal} & Classical & 0.50 & 1.00 & 1.00\\
 & ROAMS-FUT & \multirow{2}{*}{0.43} & 0.70 & 0.02\\
 & ROAMS-Kalman &  & 1.56 & 2.61\\
\midrule
\multirow{3}{*}{Blue whale} & Classical & 0.79 & 1.00 & 1.00\\
 & ROAMS-FUT & \multirow{2}{*}{0.43} & 0.67 & 0.41\\
 & ROAMS-Kalman &  & 0.94 & 0.86\\
\bottomrule
\end{tabular}
\end{table*}

The MSFE and $\text{MSFE}_\text{clean}$ columns of Table \ref{tab:animal_data_performance} demonstrate the performance of the ROAMS methods relative to the classical method, where a value below one indicates better performance of ROAMS. ROAMS-Kalman and especially ROAMS-FUT perform worse than the classical method on the cat data set. This data set was free of outliers as they were collected using accurate GPS systems, so it makes sense for the classical method to perform better here. On the polar bear, seal and whale data sets, ROAMS-FUT performs well, and much better than the classical method. 
These three data sets were collected using the Argos Doppler shift technology, which is much more prone to creating outliers compared to GPS technology.
Especially on the polar bear data under the $\text{MSFE}_\text{clean}$ metric, it performs 100$\times$better than the classical method. This strong performance is due to the presence of an extreme outlier in the testing set of the polar bear data. The seal and whale data sets similarly contain some distant outliers, hence the superior performance of ROAMS-FUT over the classical method. 

\subsection{Blue whale location tracking} \label{sec:blue_whale}
We now zoom into the performance of ROAMS to track the location of the blue whale.

\begin{figure}[h]
    \centering
    \includegraphics[width=1\linewidth]{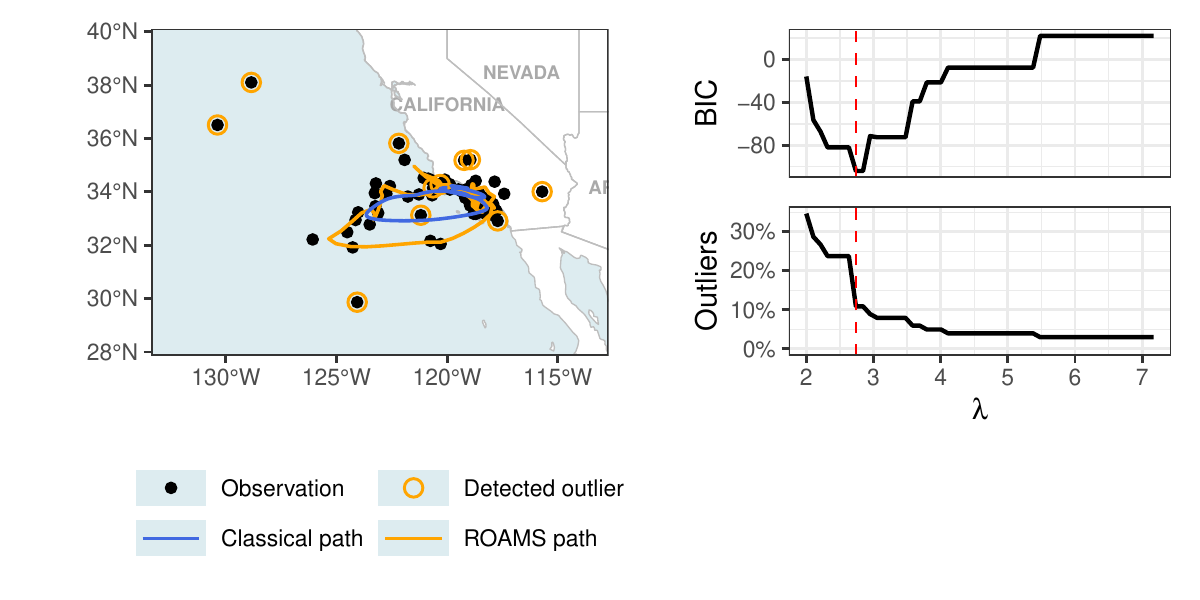}
    \caption{\textit{Left}: Map of whale location measurements over 61 days. The smoothed fit of the whale's trajectory assuming the DCRW state-space model as estimated by ROAMS and by the classical method are shown. \textit{Right}: BIC and percent outlier curves of ROAMS across 50 equally-spaced values of $\lambda$. The choice of $\lambda$ that minimises the BIC is shown by the dashed red line ($\lambda = 2.74$). This $\lambda$ detects 10.9\% of points as outliers, and is the $\lambda$ used by the ROAMS fit shown on the map on the left.}
    \label{fig:whale_map}
\end{figure}

Using ROAMS to robustly estimate the DCRW model on the full whale data set, 
Figure \ref{fig:whale_map}, left panel, shows the map of whale location measurements and the smoothed trajectory of ROAMS and the classical method. 
The smoothed trajectory is computed using the Rauch-Tung-Striebel (RTS) smoother \citep{rauch1965} based on the filtered states and parameter estimates obtained from either ROAMS or the classical method. The smoothed trajectory is plotted instead of the filtered trajectory because it provides a better estimate of in-sample states.

The right panel shows the BIC curve and percentage of detected outliers across the different values of $\lambda$; the optimal $\lambda = 2.74$ flags $\approx 10.9\%$ of non-missing timepoints as outlying, i.e. 11 out of 101 timepoints. 
A few points that are well inside the state of California were successfully flagged as outliers. A few points that are closer to the coast but still inland were not flagged, so the smoothed trajectory from ROAMS crosses into California since we did not supply any information about the geography; we leave the possible incorporation of such additional information for future research.

\begin{table*}[t]
    \centering
    \caption{Estimates of DCRW model parameters from ROAMS and from the classical method.}
    \label{tab:whale_par_est}
\begin{tabular}{cccccc}
\toprule
 & $\hat\phi$ & $\hat{\sigma}^2_{\bv,1}$ & $\hat{\sigma}^2_{\bv,2}$ & $\hat{\sigma}^2_{\bw,1}$ &
$\hat{\sigma}^2_{\bw,2}$ \\
\midrule 
ROAMS & 0.438 & 0.079 & 0.013 & 0.101 & 0.027 \\
Classical & 0.712 & 2.783 & 0.613 & 0.032 & 0.002 \\
\bottomrule
\end{tabular}
\end{table*}

The DCRW model has a total of five parameters to estimate: $\btheta = \left[\phi, \sigma^2_{\bv,1}, \sigma^2_{\bv,2}, \sigma^2_{\bw,1}, \sigma^2_{\bw,2}\right]$. Table \ref{tab:whale_par_est} compares the parameter estimates between ROAMS and the classical method. The key difference between the two methods is their estimates of the observation error variance parameters $\sigma^2_{\bv,1}$ and $\sigma^2_{\bv,2}$. The classical estimates are much larger than those of ROAMS, suggesting that the satellite system for measuring locations is very inaccurate. The estimates from ROAMS, on the other hand, indicate that the satellite system is not so inaccurate, but that occasionally it produces additive outliers. These outliers are flagged and forced to have no influence on the estimation of $\btheta$. 

In fact, the classical method attributes most of the error variance to satellite noise --- so much so that it does not attribute enough error variance to the natural randomness of the whale's movements, resulting in very small state error variances $\sigma^2_{\bw,1}$ and $\sigma^2_{\bw,2}$ when compared to those from ROAMS. Notice that the classical smoothed trajectory in Figure \ref{fig:whale_map} forms a seemingly perfect elliptical shape, reflecting the small state-to-observation-error-variance ratio. In contrast, the smoothed trajectory from ROAMS is much more willing to be `pulled around' (influenced) by non-outlying timepoints.

\section{Conclusion} \label{sec:conclusion}

We propose ROAMS, a robust estimation method for linear Gaussian state-space models that detects and mitigates the influence of observational outliers. ROAMS extends the observation equation by introducing a shift parameter at each timepoint within a mean-shift framework.
We then augment the resulting loss function with an $L_0$ penalty on the shift parameters to jointly estimate the parameters of the SSM and detect outlying timepoints.   
Computationally, the method proceeds iteratively, alternating between estimating the SSM parameters and updating the shift parameters via a hard-thresholding rule.

A key strength of ROAMS lies in its ability to separate observation noise from additive outliers. 
Classical estimation methods often conflate these two sources of error, leading to inflated observation error variances and correspondingly underestimated state error variances. ROAMS, by contrast, protects the parameter estimates from the influence of additive outliers and thus provides a more accurate decomposition of uncertainty in state-space models. The application to animal tracking for four diverse animals highlights this advantage, with our robust procedure yielding markedly lower observation error variance estimates and a more plausible smoothed trajectory compared to the classical approach for those animals where outliers in the GPS signals are clearly present.

Through a comprehensive simulation study, we demonstrate that ROAMS performs well across a variety of contamination configurations including fixed distance outliers, multi-level outliers, and clustered outliers. 
It is competitive with or superior to benchmark robust methods based on Huber and trimmed likelihoods, and closely tracks the performance of an oracle estimator that knows the true outlier locations. 
Importantly, ROAMS maintained high sensitivity and specificity for outlier detection, and its parameter estimates exhibit low root-mean-squared error (RMSE), particularly for observation error variance parameters where classical methods tend to fail. 
Increasing the outlier distance and contamination rate under the fixed distance configuration did not hinder ROAMS based on out-of-sample forecasting performance.

We also proposed a fast-updating threshold filter for online (out-of-sample) state estimation that can be used after SSM parameters have been estimated in-sample with our robust procedure. 
This filter mitigates a known weakness of traditional threshold filters; namely the cascade effect, where a single outlier detected triggers a string of false positives. 
By inflating the filtered state variance more rapidly when an outlier is detected, our modified filter is better able to recover from detection errors and improve real-time robustness against isolated outliers.

In Sections \ref{sec:study1} and \ref{sec:study2}, the FUT filter using ROAMS-estimated parameters (denoted ROAMS-FUT) had a higher out-of-sample MSFE than the Kalman filter using the same parameters (denoted ROAMS-Kalman) because the out-of-sample data was free of outliers. 
However, in Sections \ref{sec:study3} and \ref{sec:application}, ROAMS-FUT performed better than ROAMS-Kalman due to the presence of outliers in the out-of-sample data.

ROAMS uses an $L_0$ penalty to completely remove the influence of outliers. 
The $L_0$ penalty offers better protection against masking and swamping effects than the $L_1$ penalty in linear regression contexts \citep{she_outlier_2011}. 
If one is more interested in down-weighting the influence of outliers instead of completely removing them, an $L_1$ penalised variant of ROAMS where the hard-thresholding rule in \eqref{eq:thresholding_rule} is replaced with a soft-thresholding rule could be a possible avenue of future research.

ROAMS is not without limitations. State-space model estimation procedures can be quite sensitive to initial values and can often result in convergence errors if these values are supplied without thought. 
To overcome this challenge in our specific example of the first-differenced correlated random walk (DCRW) model, we used median absolute deviation of differences between observations to initialise the estimation procedure.
Furthermore, ROAMS currently assumes that outliers occur only in the observation process; it thereby
offers a simple yet effective solution to the challenge of fitting state-space models in the presence of additive outliers.
ROAMS combines strong empirical performance with useful interpretability, and it therefore holds promise as a tool for robust time series analysis across other more complex forms of contamination; which we believe to be an interesting avenue of future research.

\section*{Acknowledgements}
We thank the editor, associate editor and referee for their constructive comments which substantially improved the quality of the manuscript. 
Rajan Shankar is supported by the Australian Government Research Training Program (RTP) Scholarship.
Ines Wilms is supported by a grant from the Dutch Research Council (NWO), research program Vidi under grant number VI.Vidi.211.032.
Garth Tarr is supported by the Australian Research Council Discovery Project grants DP210100521 and DP260100348.
\newpage
\bibliography{bibliography}

\end{document}